# High-intensity wave vortices in the beam-optics domain: 3D spatial structure, single-plane models, and their propagation properties

**I. MOKHUN,* O. ANGELSKY,† A. BEKSHAEV,† Y. GALUSHKO,† AND Y. VIKTOROVSKAYA†**

*Yuriy Fedkovych Chernivtsi National University, 2, Kotsiubynskoho St., Chernivtsi 58002, Ukraine*
**i.mokhun@chnu.edu.ua*

**Abstract:** Usually, vortices in wave fields are coupled with the amplitude zeros. In contrast, the recently described high-intensity (type-II) wave vortices appear near the amplitude poles, and can only exist in a not-simply-connected space [Newton 1(3) 100060 (2025)]. We try to expand the concept of type-II optical vortex (OV) to paraxial beam-like optical fields. It is shown that the "true" type-II OV-fields can only exist in a multiply-connected 3D space where the spatial "holes", impenetrable for the electromagnetic field, are formed by the purposefully created threadlike material inclusions. However, the transverse cross section of a "true" type-II OV can be reproduced in a certain plane of the usual "free" space where the hole is signified by a zero-amplitude area. In this single plane, the planar model of any type-II OV-field can be realized as well as explored theoretically and in experiments; the results are compared with the known properties of conventional vortex fields. Upon transition to the 3D space (model-field propagation), its type-II nature is expectedly destroyed. However, in the propagation process, interesting and potentially applicable "bottle" structures with high concentration of the optical energy and angular momentum appear. Methods for practical generation of the discussed field structures are proposed, including those involving the programmable light modulators and computer-synthesized holograms. Results of computer modeling show reasonable agreement with the experimental data.

## 1. Introduction

Vortices are specific structures that occur in various physical systems. The patterns of their emergence, scale and characteristic behavior differ radically depending on the processes and physical environments in which they are realized. The vortices and vortex-like structures penetrate the whole physical world, from cosmological foundations of Universe, through turbulent formations in continuous media, up to atoms and elementary particles [1,2]. A peculiar attention is attracted to the vortex phenomena in wave processes which are ubiquitous in physical world and demonstrate the remarkable unity of the laws controlling the most fundamental dynamical and geometrical features of physical systems [1–10].

In this view, the optical waves are of special interest not only 'per se' but also as a productive model of physical waves of arbitrary origination, from cosmological perturbations and ocean tides to quantum entanglement and vacuum excitations. Optical vortices (OV) are in the focus of research efforts during past decades (see, for example, [3–11]), and their studies have essentially expanded the general understanding and specific knowledges of the vortex phenomena. An OV is formed near a specific point (line in 3D) – "OV core" where the wave phase is indeterminate (singular). As a result, the wavefront in the OV region obtains the specific helical shape with the screw pitch equal to $S\lambda$ (phase increment $2\pi S$) where $\lambda$ is the wavelength, and $S$ is an integer number (topological charge) [3,6–10]. Due to the helical phase, a specific circular energy flow occurs in the vortex region, being the source of the so-called orbital angular momentum (OAM) [11–14]. Importantly, the OAM emerges from the peculiar spatial structure of the wave, and may exist even in scalar waves, which differs it from another type of optical angular momentum, spin angular momentum, that arises from elliptical (circular) wave polarization [11–16]. An important characteristic feature of the OVs is that the wave amplitude vanishes at the singular point [6–12].

Today, the unique physical characteristics of the OVs have been well studied and have formed the basis for multiple promising applications: optical manipulation and diagnostics [17,18,19], optical communication systems [20,21], super-resolution microscopy [22,23], astronomical instruments [24,25], and other practically valuable methods and devices (see, e.g., Refs.[6,7,26]).

In this context, a great interest is excited by the recent information about conceptually new singularities in scalar wave fields, the so-called type-II vortices [27,28]. While in the conventional OV-field (type-I OV) with the topological charge $S$, the wave amplitude depends on the off-core radius $\rho$ as $\propto \rho^{|S|}$ (i.e. tends to zero at the core $\rho = 0$), the situation for the type-II vortices is quite opposite. At the type-II vortex core, the wave amplitude behaves as $\propto \rho^{-|S|}$: instead of the usual zero, the amplitude pole takes place [27,28]. In this case, the phase singularity (indeterminacy) at $\rho = 0$ is coupled with the singular intensity (theoretically infinite value at $\rho = 0$), which is physically forbidden. Accordingly, the type-II vortices may only appear in not simply-connected spatial regions, e.g., of the toroidal topology, where the nearest vicinity of the vortex core is not available for the physical field [27,28]. Remarkably, the size of this "unavailable" area may be arbitrary small, up to the deep subwavelength scale, such that the "external" vortex behavior is practically independent of it.

As shown in [27,28], the type-II vortices are rather generic and may appear in random wave fields localized in a multiply connected 2D space. The probability of their spontaneous generation as well as of the existence of material inclusions acting as physically "excluded" regions ("holes"), forbidden for the waves, varies from 50% to almost 100% under certain additional field-formation conditions. For the very characteristic example of fields formed by ocean tidal waves, the holes are naturally formed by islands, while the vortex-generated factor is the so-called Coriolis force playing a substantial role in many geophysical processes.

A simple theory of monochromatic waves shows that the wave structures with type-II vortices satisfy the Helmholtz equation, and can be realized in various physical domains, including optics [27,28]. In more detail, the characteristics of such optical formations, as well as possible methods of their generation and control, are considered with the examples of surface evanescent waves. There, the excluded (physically unavailable) areas may be created by either real holes in a surface supporting the waves, or by nanoparticles (defects) introduced in a homogeneous surface.

The existing analysis and experimental realizations of the type-II OVs [27,28] essentially involve the two basic aspects:

- the optical perturbations are of 2D character, i.e. are not propagating optical beams which form the most common model of optical fields, widely used in theory and applications;
- the excluded regions, or holes, realizing the multiply-connected field-accommodating space, are of subwavelength sizes, thus enabling the almost ideal type-II OV structures.

Both these aspects are physically consistent and practically available (although with certain precautions) in optical domain. However, none of them is compatible with the conventional concept of propagating light beam. In this paper, we try to expand the type-II OV ideas to the usual domain of beam optics. Simultaneously, the specific properties and manifestations of vortex fields that are important in this connection, are inspected and analyzed.

Herewith, the following ways of development will be in the spotlight:

(i) Possible implementation of propagating type-II OV fields in the not simply connected 3D space. To this purpose, possible realizations of the 3D holes should be considered as specific material inclusions with sharply distinct electric and/or magnetic properties making them physically unavailable for electromagnetic field. In this approach, the "prototype" 2D type-II OV is considered as a transverse $(x, y)$ cross section of the beam propagating along the longitudinal $z$-axis, and the hole forms a cylindrical volume enclosing the near-axis region. The existence and properties of such "hole-supported" fields is of great interest.

(ii) Obviously, it is difficult to create a set of 3D material "holes" supporting the desirable combination of 3D type-II OV fields. At the same time, it seems quite possible that the 2D type-II OV structures can be created in a single specially assigned plane, using traditional spatial light modulators (SLM) or synthesized hologram techniques [29,30]. The details, regulations and limitations of the corresponding approaches need to be evaluated and analyzed.

(iii) While the 2D type-II OV structures can be generated in a single plane, the properties of such fields and patterns of their formation can be analyzed using methods traditional for the singular optics. The networks of interrelated and interacting singularities (singular skeletons) appear, and the questions arise about the regularities they obey, and how do these regularities correlate with the known ones "governing" the behavior of traditional 2D fields with type-I OVs? Can the type-II OV networks play the role of skeletons, qualitatively determining the behavior of other field parameters at any point, as the type-I OVs play? And is it possible to assign a desirable in-plane polarization as well as to combine differently-polarized 2D fields producing vectorial type-II OV structures?

(iv) And, finally, the "artificial" type-II OV fields, created in a single plane, will inevitably propagate along the $z$-direction (at least under the paraxial-optics conditions which are typical for the beam optics, and are supposed to fulfil in the main contents of this paper). In this context, it is unclear how the structure of these "single-plane" type-II OV-fields is transformed as the field propagates in free space, and which specific field configurations and useful properties can be revealed in this process?

In this paper we try to answer these and relating questions.

## 2. Propagating type-II optical vortex

We start with a simple theoretical example illustrating the existence and physical nature of propagating 3D fields with the type-II OV structure. As usual, we deal with monochromatic fields with the angular frequency $\omega$, for which the time-dependent electric and magnetic vectors can be presented via the complex-valued phasors as $\{\boldsymbol{E}(t), \boldsymbol{B}(t)\} = \mathrm{Re}(\{\mathbf{E}, \mathbf{B}\}e^{-i\omega t})$ [31]. In this representation, Maxwell equations for the electromagnetic field in free space have the form:

$$\nabla \cdot \mathbf{E} = 0, \quad \mathbf{E} = -\frac{1}{ik}\nabla \times \mathbf{B}\,, \tag{1}$$

$$\nabla \cdot \mathbf{B} = 0, \quad \mathbf{B} = \frac{1}{ik}\nabla \times \mathbf{E}\,, \tag{2}$$

$k = \omega/c$ being the wavenumber, $c$ the light velocity. Like in 2D situations, the type-II OV can only exist in a not simply-connected space, such that the OV core and its nearest vicinity are excluded from the field-accommodating space. Let the field propagate along the longitudinal axis $z$; then, the $(x, y)$-plane is its transverse cross section, and we suppose that the OV core is situated at $(x = 0, y = 0)$. In this geometry, a 'hole' is a 3D body enclosing the axis of propagation $z$. For simplicity, we suppose that the hole is of the cylindrical shape with the radius $a$ (see Fig. 1), and introduce the cylindrical coordinates $\rho = \sqrt{x^2 + y^2}$, $\varphi = \arctan(y/x)$. In such a configuration, the type-II OV with the topological charge $S$ is described by relation [27,28]

$$u_\alpha(\rho,\varphi) = A_\alpha(\rho)\exp[i\Phi(\varphi)] = \frac{K_\alpha}{\rho^{|S|}} e^{iS\varphi} \tag{3}$$

where $u_\alpha$ denotes any component of the electric or magnetic vectors entering Eqs. (1) and (2), $A_\alpha$ and $\Phi$ are its amplitude and phase, $K_\alpha$ is the corresponding normalization constant, and we suppose that Eq. (3) holds at the initial plane $z = 0$. The relative permittivity and permeability of the external (outside the hole) space are $\mu_{out} = \varepsilon_{out} = 1$, which corresponds to vacuum or air. Properties of the electromagnetic field and of the hole material should be adjusted such that the electric and magnetic field cannot penetrate into the region $\rho < a$.

Our immediate task is to find the solution to Eqs. (1), (2), which is of the form (3) and describes the propagating beam in the geometry outlined by Fig. 1. To this end, we suppose that there is no bound charge and current densities [32] induced near the cylinder surface; then, the boundary conditions [32] at $\rho = a$ dictate the continuity of the normal component of the electric displacement, $\varepsilon_{out}\mathbf{E}_\perp = \mathbf{E}_\perp = \varepsilon_{in}\mathbf{E}_\perp^{in}$, and of the tangential component of the magnetic intensity, $\mu_{out}^{-1}\mathbf{B}_\parallel = \mathbf{B}_\parallel = \mu_{in}^{-1}\mathbf{B}_\parallel^{in}$ [32]. This means that the normal electric and tangential magnetic fields inside the cylinder vanish, $\mathbf{E}_\perp^{in} = 0$ and $\mathbf{B}_\parallel^{in} = 0$, if

$$\mu_{in}^{-1} = \infty,\ \varepsilon_{in}^{-1} = 0. \tag{4}$$

In turn, the tangential electric and normal magnetic fields will be eliminated if the corresponding components are absent in the external field. For this reason, we look for the solution to Eqs. (1), (2) in the form in which the electric field is exactly normal ($E_\varphi = E_z = 0$) while the magnetic field is exactly tangential ($B_\rho = 0$). Namely,

$$E_\rho = K\frac{e^{i\sigma\varphi}}{\rho}e^{ikz}, \qquad E_\varphi = E_z \equiv 0, \tag{5}$$

$$B_\rho \equiv 0, \qquad B_\varphi = K\frac{e^{i\sigma\varphi}}{\rho}e^{ikz}, \qquad B_z = -K\frac{\sigma}{k\rho}\frac{e^{i\sigma\varphi}}{\rho}e^{ikz} \tag{6}$$

where $K$ is the normalization constant and $\sigma = \pm 1$ denotes the vortex sign. The solution form (5), (6) restricts our analysis to the generic case where the absolute topological charge of the OV equals to $|S| = 1$. This is made not only for simplicity but is also dictated by the general rule that only single-charged type-I vortices are stable and propagate with preserving their topological nature [6–10]. In this Section, we follow the similar reasoning for the type-II OVs, and exclude the case $|S| > 1$ from the present study.

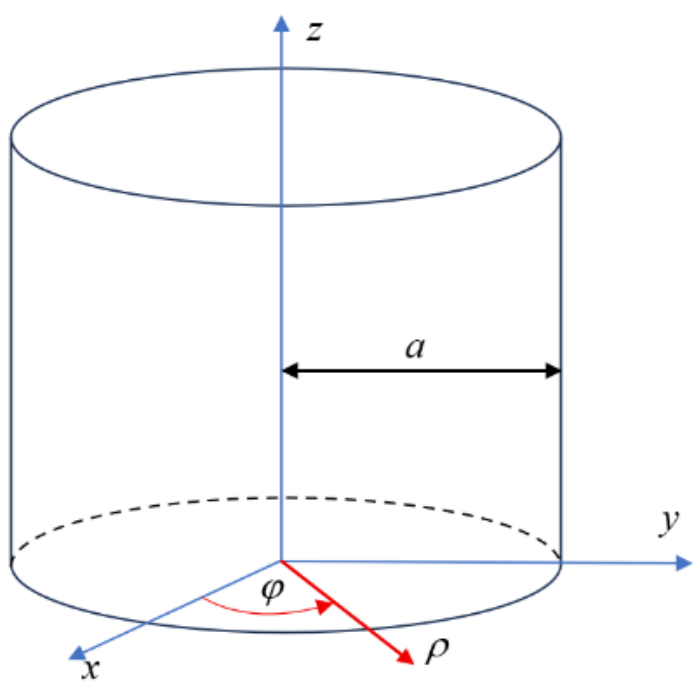

Fig. 1. Cylindrical hole supporting the type-II optical vortex (5), (6), and accompanying coordinate frames.

Substituting Eqs. (5) and (6) into (1), (2), one finds:

$$\nabla\cdot\mathbf{E}=\frac{1}{\rho}\frac{\partial(\rho E_\rho)}{\partial\rho}+\frac{1}{\rho}\frac{\partial E_\varphi}{\partial\varphi}+\frac{\partial E_z}{\partial z}=0,\;\;\nabla\cdot\mathbf{B}=0,$$

$$\frac{1}{ik}\nabla\times\mathbf{E}=\frac{1}{ik}\left[\left(\frac{1}{\rho}\frac{\partial E_z}{\partial\varphi}-\frac{\partial E_\varphi}{\partial z}\right)\mathbf{e}_\rho+\left(\frac{\partial E_\rho}{\partial z}-\frac{\partial E_z}{\partial\rho}\right)\mathbf{e}_\varphi+\frac{1}{\rho}\left(\frac{\partial(\rho E_\varphi)}{\partial\rho}-\frac{\partial E_\rho}{\partial\varphi}\right)\mathbf{e}_z\right]$$
$$=B_\varphi\mathbf{e}_\varphi+B_z\mathbf{e}_z$$

($\mathbf{e}_{\rho,\varphi,z}$ is the unit vector along the corresponding coordinate axis); that is, Eqs. (2) and the 1st Eq. (1) are satisfied. However, substitution of expressions (5) and (6) into the 2nd Eq. (1) yields

$$-\frac{1}{ik}\nabla\times\mathbf{B}=\left(1+\frac{1}{(k\rho)^2}\right)E_\rho\mathbf{e}_\rho-\frac{2i\sigma}{(k\rho)^2}E_\rho\mathbf{e}_\varphi \tag{7}$$

which differs from the correct result $E_\rho\mathbf{e}_\rho$ required by (1) solely by summands of the relative order $(k\rho)^{-2}$, negligibly small in the paraxial case,

$$k\rho>ka\gg 1. \tag{8}$$

This result leads to the conclusion that Eqs. (5), (6) realize, although in the paraxial approximation, a possible 3D generalization of a beam with type-II OV, which propagates along the longitudinal *z*-direction. Remarkably, the propagation is diffraction-free and, theoretically, lossless: the beam's transverse shape and local intensity does not depend on the propagation distance *z*.

A few words should be said in relation to the specific medium with the parameters (4) that enables such a propagation. The requirements (4) comply with properties of the perfect-metal (perfect-conductor) medium, where electric permittivity $\varepsilon$ approaches negative infinity, $\varepsilon=-\infty$ , or has an infinitely large imaginary component, while the relative magnetic permeability is exactly zero, $\mu=0$, and the electric and magnetic fields cannot exist [32]. On the other hand, the properties (4) may characterize a specific "artificial dielectric". Anyway, a medium, necessary for formation of a genuine hole for the electromagnetic field in the region $\rho<a$, can be realized in practice with some limitations and approximations. The main limitation is that the charge carriers in a perfect conductor should move "without inertia", which implies not very high frequency (~ percents of the plasma frequency) [32], and apparently contradicts to the paraxial condition (8) requiring the wavelength to be as small as possible. Nevertheless, the necessary conditions for $\varepsilon$ and $\mu$ can be achieved, at least approximately, which means that not only 2D type-II vortices described in [27,28] exist, but their propagating 3D analogs are possible. Their propagation is supported by the 'hole' cylinder in Fig. 1, "forbidden" for the electromagnetic field penetration; the propagation is closer to diffraction-free and lossless, the less the electromagnetic field penetrates into the cylinder.

To finalize this Section, we present the dynamical characteristics [9] of the type-II field (6): energy density $W=\frac{1}{16\pi}(|\mathbf{E}|^2+|\mathbf{B}|^2)$ and the Poynting vector $\mathbf{F}=c^2\mathbf{P}=\frac{c}{8\pi}\mathrm{Re}(\mathbf{E}*\times\mathbf{B})$ (energy flow density proportional to electromagnetic momentum density $\mathbf{P}$ coinciding in this case with the orbital, or canonical, momentum density $\mathbf{P}_\mathrm{O}$):

$$W=\frac{1}{8\pi}\frac{|K|^2}{\rho^2},\;\;\mathbf{P}=\mathbf{P}_\mathrm{O}=P_z\mathbf{e}_z+P_\varphi\mathbf{e}_\varphi=\frac{1}{8\pi c}\frac{|K|^2}{\rho^2}\left(\mathbf{e}_z+\frac{\sigma}{k\rho}\mathbf{e}_\varphi\right), \tag{9}$$

and the OAM density along the longitudinal axis

$$\mathbf{L}_z = L_z \mathbf{e}_z = \rho \mathbf{e}_\rho \times P_\varphi \mathbf{e}_\varphi = \frac{\sigma}{8\pi\omega} \frac{|K|^2}{\rho^2} \mathbf{e}_z \,. \tag{10}$$

Equations (9), (10) are valid in the paraxial approximation (8), i.e. upon neglecting summands of the relative order $(ka)^{-2}$. Note that the radial dependence of the light energy $W$, longitudinal momentum $P_z$ and the longitudinal OAM $L_z$ obey the universal inverse-power law $\propto \rho^{-2}$ [27,28] with the sharp maximum exactly at the hole boundary (Fig. 2).

The type-II OV field, characterized by Eqs. (5), (6), (9), (10) and illustrated by Figs. 1, 2, represent a simplest idealized example. Expectably, if there exist several material inclusions with appropriate electric and magnetic properties, making them impenetrable for electromagnetic field, and the inclusions are not obligatory cylindrical in shape but form a set of 3D "threads of darkness" (3D holes) elongated in the *z*-direction, similar solutions to the Maxwell equations may be found. Near the hole surfaces, such solutions would correspond to the type-II OV fields with the topological structure similar to that of Eqs. (5), (6), propagating along the spatially variable curvilinear holes. In general, the whole field pattern would resemble the situation of "knotted and tangled threads of darkness" spontaneously emerging in freely propagating light fields containing networks of conventional type-I OVs [33].

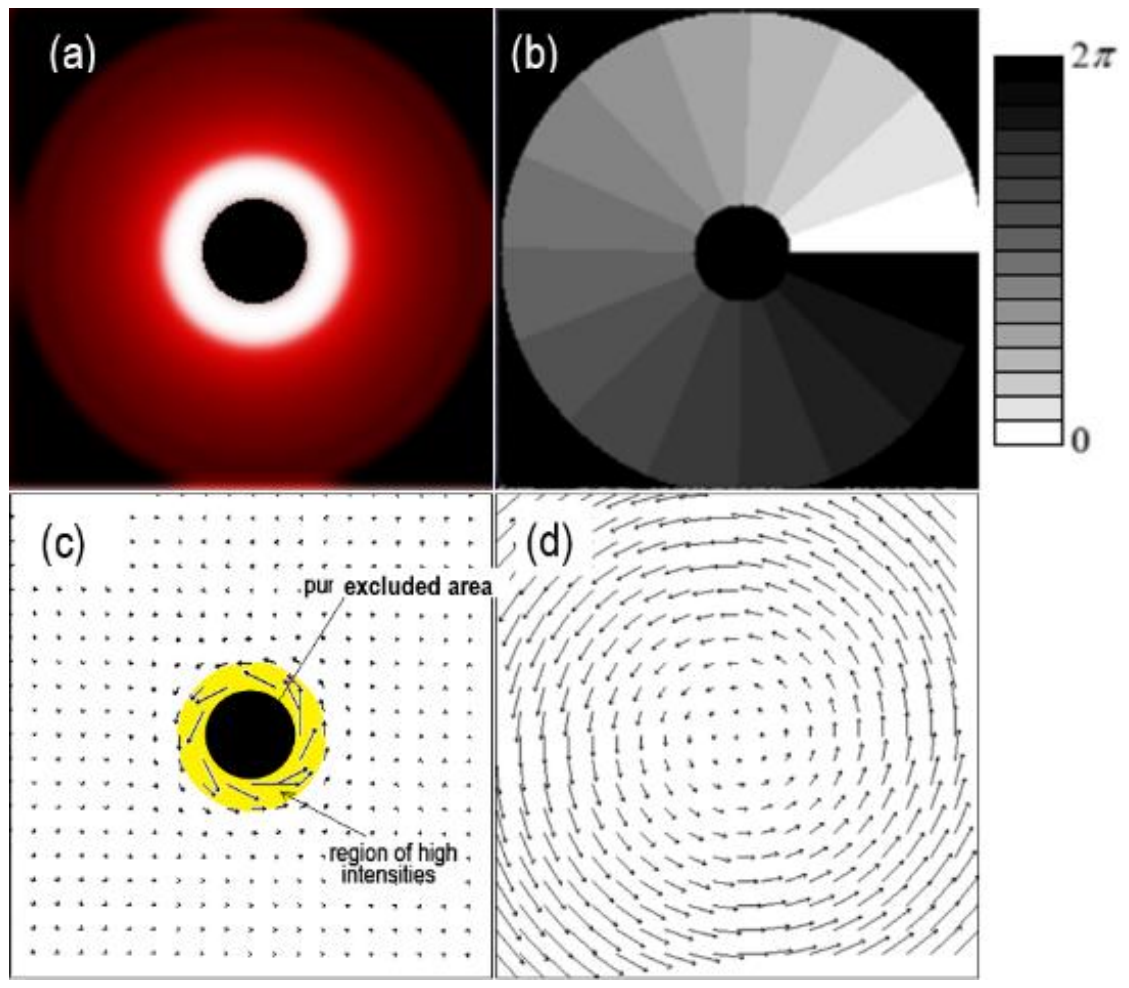

Fig. 2. Transverse distribution of the dynamical characteristics in the type-II OV field (5), (6): (a) field intensity (energy density, longitudinal momentum, OAM, see Eqs. (9), (10)); (b) phase (helical surface, see Eqs. (3) and (5), (6)); (c) transverse Poynting vector component (2nd term of Eq. (9)), arrows show the component direction, their lengths are proportional to its magnitude. The dark spot in the center corresponds to the cylindrical hole of Fig. 1 where the field does not exist. Panel (d) shows the transverse Poynting vector component in the "usual" type-I OV field (for comparison).

## 3. Singularity network of the type-II vortices in comparison with the conventional type-I fields

The example of the previous Section illustrates the principles underlying the 3D type-II OVs. Our next problem originates from the transparent analogy with the conventional type-I OVs. As is known [6–10], multiple type-I OVs spontaneously appear in paraxial optical fields, even under stochastic conditions (speckle fields); the separate singularities interact, are mutually interconnected and obey regular rules in their localization and interrelation thus forming the singularity network ("singular skeleton") of the field [7,10,34–41]. Accordingly, an interesting question appears about the interdependence and correlations between multiple type-II singularities. The above example shows that spontaneous presence of multiple type-II OVs within a conventional optical field, spreading across a homogeneous space, is improbable. Their creation requires special preparation of multiple 3D holes – material inclusions with the specific electromagnetic properties (e.g., characterized by Eq. (4)), prohibiting the field penetration ("islands" in the electromagnetic "ocean" [27,28]). In principle, such configurations are practically available but require special efforts for their creation and support so that their practical relevance is problematic.

However, arbitrary combinations of type-II OVs, including stochastic ones, can be formed in a single plane without the special material inclusions. A general way to create such fields starts with the conventional "generic" scalar paraxial light field characterized by the complex amplitude distribution [9,12]

$$U(x,y) = A(x,y)e^{i\Phi(x,y)} \tag{11}$$

where $A(x,y)$ is the local wave amplitude and $\Phi(x,y)$ is its phase. Then, it is possible to fabricate a transparency with the amplitude transmission $\propto [A(x,y)]^{-2}$ (despite the essential practical difficulties, this operation is not forbidden in principle), and after an input "source" wave (11) passes this transparency, the "inverse-amplitude" field with the complex amplitude distribution

$$V(x,y) = \frac{C}{A(x,y)}e^{i\Phi(x,y)} \tag{12}$$

is formed ($C$ is a certain normalization constant). If the "source" field (11) is the speckle field with multiple type-I OVs, the resulting field (12) is also a stochastic field but it contains a singular skeleton formed by the type-II OVs. Of course, these are not "true" type-II OVs, and exist only in a single plane, but, restricting to this plane and inspecting their properties, it would be convenient to preserve their name "type-II OV". Also for convenience, the source field (11) will be referred to as "type-I field" whereas the artificially generated field (12) – as "type-II field".

At the first stage, let us consider a single type-II OV associated with the field (12). For simplicity we suppose that its core is localized at the coordinate origin ($x = 0, y = 0$), and the OV is isotropic, i.e. the near-core amplitude depends solely on the off-core distance $r$ [9,34]. Then, the near-core expression of the field (12) can be presented in the form

$$V_a(x,y) = [1 - Q_a(x,y)]\frac{C_a}{\rho}e^{i\sigma\varphi} \tag{13}$$

where

$$Q_a(x,y) \equiv Q_a(\rho) = \begin{cases} 1, \sqrt{x^2+y^2} \le a\,; \\ 0, \sqrt{x^2+y^2} > a\,. \end{cases} \tag{14}$$

The multiplier $1 - Q_a(x,y)$ is necessary because no physical amplitude, nor physical phase gradient can reach infinite values formally required by Eq. (12) at the OV core. Actually, it creates the zero-amplitude OV-core vicinity (see the cross section of the field-impenetrable cylinder in Fig. 1), which represents the "trace" of the hole realizing the multiply connected space indispensable for the type-II vortex.

Similarly to Eqs. (9), (10) and based on the known results for the Poynting vector of a paraxial optical field [7–14], the momentum components near the core of a scalar type-II OV (13), (14) can be presented as

$$P_\rho = 0\,;\ \ P_\varphi = \frac{|C_a|^2}{8\pi\omega}[1 - Q_a(\rho)]\frac{\sigma}{\rho^3}\,;\ \ P_z = \frac{|C_a|^2}{8\pi c}[1 - Q_a(\rho)]\frac{1}{\rho^2} \tag{15}$$

while the OAM density of this field is

$$L_z = \frac{|C_a|^2}{8\pi\omega}[1 - Q_a(\rho)]\frac{\sigma}{\rho^2}. \quad (16)$$

Confronting with Eqs. (9) and (10) shows that, in terms of dynamical characteristics, the simple scalar model of the type-II OV based on Eqs. (13) and (14) turns out to be quite equivalent to the “true” vector vortex field of Eqs. (5), (6).

Now we apply the methodology of Eqs. (11) – (14) to the type-II analogue of a generic random field which can be obtained via the transformation (12). Its complex amplitude can be written in the form

$$V(x,y) = \frac{1}{N}\sum_{l=1}^{N}[1 - Q_a(x - x_l, y - y_l)]\frac{C}{A(x,y)}e^{i\Phi(x,y)} \quad (17)$$

where $A(x,y)$ and $\Phi(x,y)$ describe the source speckle field of the form (11) with $N$ type-I singularities localized at points $x_l, y_l$ of the considered transverse plane.

An important feature of the field (17) is that its phase distribution $\Phi(x,y)$ coincides with the phase of the source speckle field. Accordingly, a lot of regularities, characteristic for the conventional scalar vortex field, can be expanded to the fields (17) containing the type-II OVs. Moreover, this statement can be applied even for non-uniformly polarized waves whose orthogonal components are of the form (17), since it is valid for each of the components.

In particular, like in conventional fields [6–10], the singularities of the type-II field (17) are not independent but form a coherent and interrelated singular network, which can be considered as a certain skeleton for the amplitude and phase spatial structure. Moreover,

1. By analyzing the distributions of the field characteristics in the region immediately adjacent to the singularity, it is possible to predict the “overall” behavior of the field parameters (at least qualitatively, at the probabilistic level) at any point in the field.
2. Since the singular networks of different parameters (amplitude, phase, polarization, energy flows, etc. [12]) are interconnected [35], such analysis reveals informative relationships between the complementary characteristics of an optical field reflecting their global correlation/anticorrelation properties. Such relationships, realized between the distributions of intensity and phase, were first pointed out by I. Freund and colleagues [36,37]. In particular, it was shown that in regions where the intensity changes slowly, the phase changes relatively quickly, and vice versa.

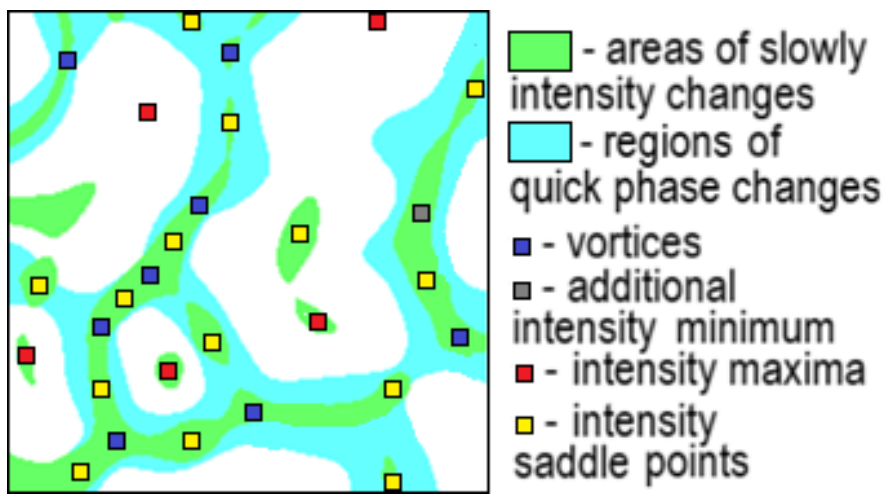

Fig. 3. “Anticorrelation” between the intensity and phase in a scalar speckle field [10]. Red squares denote the intensity maxima; black square shows the non-singular intensity minimum. Other notations are explained in the figure.

For example, it has been shown [10,35–40] that in a conventional type-I speckle field, the modulus of the average phase gradient at a stationary-intensity point is $\sqrt{2}$ times greater than at any other point (excluding the OV zones). More than 90% of stationary-intensity points are located in regions of rapid phase change (see, for example, Fig. 3). A similar rule holds for stationary-phase points: the modulus of the average intensity gradient at a stationary-phase point is $\sqrt{2}$ times greater than at any other point in the field. Let us show that the same regularities take place for the type-II speckle fields (17) produced via the transformation (12).

For the analysis we employ the real and imaginary parts of the complex amplitude (11):

$$U = \mathrm{Re} + i\mathrm{Im},\;\; I = \mathrm{Re}^2 + \mathrm{Im}^2,\;\; A = \sqrt{I},\;\; \Phi = \arctan\left(\frac{\mathrm{Im}}{\mathrm{Re}}\right) \tag{18}$$

($I = I(x, y)$ denotes the field intensity). Obviously, for the conventional type-I random fields, the following relations take place:

1. Mean real and imaginary parts of the complex amplitude vanish, together with the mean values of their spatial derivatives:

$$\left\{\overline{\mathrm{Re}},\;\; \overline{\mathrm{Im}},\;\; \overline{\frac{\partial \mathrm{Re}}{\partial x}},\;\; \overline{\frac{\partial \mathrm{Re}}{\partial y}},\;\; \overline{\frac{\partial \mathrm{Im}}{\partial x}},\;\; \overline{\frac{\partial \mathrm{Im}}{\partial y}}\right\} = 0 \tag{19a}$$

(overbar denotes the statistical averaging).

2. Mean squares of the real and imaginary parts equal to half the mean intensity:

$$\overline{\mathrm{Re}^2} = \overline{\mathrm{Im}^2} = \overline{A^2}/2\,. \tag{19b}$$

3. And finally, the mean values of the squares of the moduli of the real and imaginary parts' gradients are equal (because the statistics of Re and Im are the same). Let us denote this mean value as $D$:

$$\overline{|\nabla \mathrm{Re}|^2} = \overline{|\nabla \mathrm{Im}|^2} = D. \tag{19c}$$

Now, based on Eqs. (19a) – (19c), one can easily derive the simple relation involving the mean squares of the amplitude, intensity, and phase gradients $\overline{|\nabla A|^2}$, $\overline{|\nabla I|^2}$, $\overline{|\nabla \Phi|^2}$:

$$\overline{|\nabla \mathrm{Re}|^2} + \overline{|\nabla \mathrm{Im}|^2} = 2D = \overline{|\nabla A|^2} + \overline{A^2}\cdot\overline{|\nabla \Phi|^2} = \frac{1}{4\bar{I}}\overline{|\nabla I|^2} + \bar{I}\cdot\overline{|\nabla \Phi|^2}. \tag{20}$$

It entails that, on the average for the whole field,

$$\overline{|\nabla \Phi|^2} = \frac{D}{\bar{I}}\,, \tag{21a}$$

while at the stationary points of intensity, where $\nabla I = \nabla A = 0$,

$$|\nabla I|^2 = 0,\;\; \overline{|\nabla \Phi|^2} = \frac{2D}{\bar{I}}\,. \tag{21b}$$

Hence, the similar relations for the type-II field (12), (17) can be derived via replacement $A \rightarrow C/A, I \rightarrow (|C|/A)^2$. As a result, we have confirmed for both the type-I and the type-II fields, that in the areas of stationary intensity (where the amplitude varies slowly), the phase changes, on the average, $\sqrt{2}$ times faster than at other points of the field.

Analogous conclusions, concerning the intensity fluctuations, can be made for stationary points of the field phase (excluding singularities). Then, for the type-I field (11), on the average for the whole field

$$\overline{|\nabla I|^2} = 4\bar{I}D, \tag{22a}$$

while at the stationary points of the phase distribution,

$$|\nabla \Phi| = 0,\;\; \overline{|\nabla I|^2} = 8\bar{I}D. \tag{22b}$$

Again, this means that in areas where $|\nabla\Phi| \approx 0$, i.e. the phase changes slowly, the intensity of both type-I (11) and type-II (12), (17) fields changes approximately $\sqrt{2}$ times faster than at other points of the field.

In general, the main outcome of the above analysis is that for both type-I and type-II scalar fields, the same correlation regularities for the singular-skeleton formation apply. For example, the sign principle [41] is fulfilled for both type-I and type-II OVs, which qualitatively controls the topological organization of the singularity networks. In particular, it determines the laws of the singularities' combination according to which adjacent vortices have opposite signs and are connected by lines of equal phase. Opposite situations, where adjacent OVs are of the same sign, are much less common and strictly differ from the previous configurations: there are no equiphase lines uniting the adjacent OVs but a phase saddle point is formed between them (some illustrative examples are presented in Fig. 3).

## 4. Vector type-II fields. Polarization singularities and "Full Poincaré Beam"

In optical domain, scalar fields considered in the previous Section actually represent the fields with arbitrary polarization uniformly distributed across the plane of analysis. Obviously, a superposition of differently polarized type-II fields with different spatial profiles, $V_1(x,y)$ and $V_2(x,y)$, produces an inhomogeneously polarized field. Such superpositions realize vector type-II fields which may reveal rich polarization textures with associated polarization singularities. As an example, let us consider the formation of the so-called "full Poincare beam", an elementary polarization structure with a C-point (where the field is circularly polarized) surrounded by an s-contour (a line along which the field is linearly polarized) [1,10,42–44]. In the standard approach [10,44], such a beam can be organized as a result of the superposition of two waves with opposite circular polarizations: the first with an isotropic type-I OV and the second of the non-singular Gaussian profile (see Fig. 4).

For example, suppose the OV field is left-polarized, and the Gaussian one is right-polarized. At the point where the OV-field amplitude is zero (OV core), the total field is circularly right-polarized (see Figs. 4a, 4b): a C-point is formed. Accordingly, an s-contour arises where the intensities of the OV and the Gaussian field are equal. In the case of circular symmetry, presented in Fig. 4, the s-contour has a circular shape. At the same time, all states of polarization, intermediate between the circular and linear, are formed in points of the region bounded by this contour. The same states of polarization can be observed on one of the hemispheres of the Poincaré sphere; this explains the term "full Poincaré beam" [44].

A similar wave configuration emerges if a symmetric circularly polarized type-II OV is used instead of a type-I vortex field. As in the previous case, an s-contour is formed at the location where the intensities of the superposed fields are equal (see Fig. 4c and 4d). Its shape is also circular, and the circle radius can be easily estimated. If the complex amplitude in the near-core vicinity of the type-I OV is described by

$$U_1(\rho) = C_1 \rho e^{i\varphi}, \tag{23}$$

and the near-core type-II OV field is characterized by Eq. (13) with $\sigma = 1$, $\rho > a$ (both vortex fields are assumed in Fig. 4 left-polarized with a topological charge of +1), the condition for s-contour can be written in the form

$$\left\{C_1 \rho e^{i\varphi},\ \frac{C_a}{\rho} e^{i\varphi}\right\} = A_0 \exp\left(-\frac{\rho^2}{w^2}\right) \cong A_0\,. \tag{24}$$

Here, the r.h.s. represents the Gaussian beam with maximum amplitude $A_0$ and radius $w$, while the second (approximate) equality implies that $\rho \ll w$ in the near-core region. Solution of Eq. (24) gives the s-contour radius values

$$\begin{cases} \rho_{\mathrm{sI}} = \left|\dfrac{A_0}{C_1}\right|, & \text{type} - \mathrm{I} \quad \mathrm{OV}; \\ \rho_{\mathrm{sII}} = \left|\dfrac{C_a}{A_0}\right|, & \text{type} - \mathrm{II} \quad \mathrm{OV}. \end{cases} \tag{25}$$

Regarding the relations between $C_1$, $C_a$ and $A_0$, $\rho_{\mathrm{sII}}$ may be higher or smaller than $\rho_{\mathrm{sI}}$; generally, the s-contour size is not directly determined by the law of the amplitude variation near the OV core. However, a characteristic feature of the type-II superposition is that there is not a C-point but a whole C-area near the field center corresponding to the hole trace $\rho < a$ (Fig. 4c, d). Like in the type-I superposition, this area preserves the polarization of the smooth component. Additionally, the area, where the polarization state changes continuously and the Poincare hemisphere is mapped, is of the ring-like shape and situated between the hole trace and the s-contour $a < \rho < \rho_{\mathrm{sII}}$ (Fig. 4c, d). The polarization handedness in this area coincides with the OV-field handedness (left in Fig. 4), while the Gaussian-beam handedness prevails outside it (for the type-I OV, just the inverse situation occurs, Fig. 4a, b). Finally, if the hole diameter is small enough, then near its boundary $\rho \gtrsim a$ the intensity of the type-II OV component becomes significantly higher than the smooth-beam intensity, and the resulting field is practically circularly polarized, so the full Poincare hemisphere is mapped onto the ring $a < \rho < \rho_{\mathrm{sII}}$.

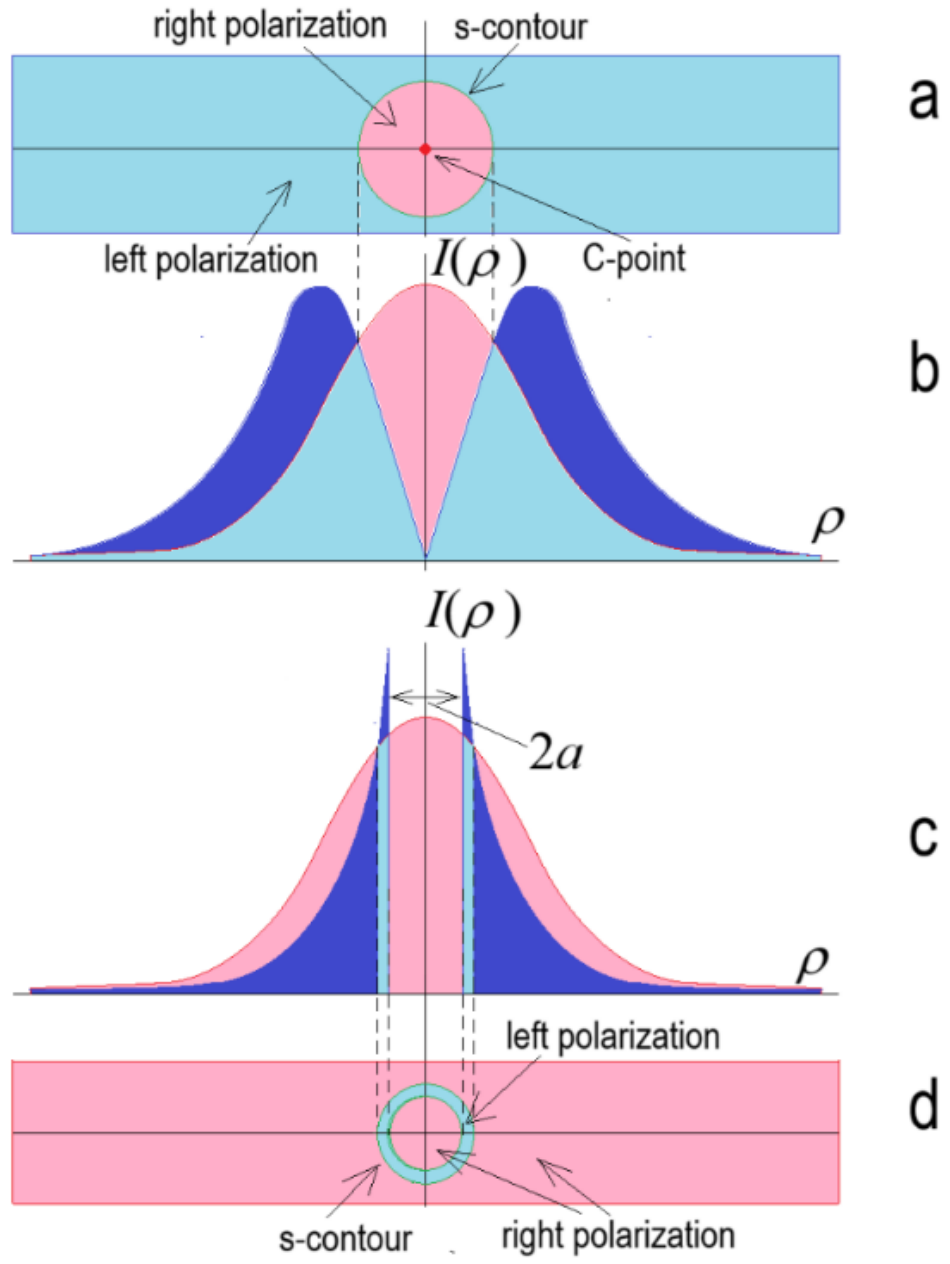

Fig. 4. Superposition of circularly polarized fields: Gaussian beam (right-hand polarization) and vortex beams of types I and II (left-hand polarization). (a, d) spatial structure of inhomogeneously polarized regions; (b, c) corresponding intensity distributions of oppositely polarized fields. Images (a, b) explain the formation of a conventional C-point and s-contour as a result of the superposition of a Gaussian beam and an isotropic type-I OV; (c, d) show the formation of an inhomogeneously polarized field with the C-area and s-contour due to superposition of a Gaussian beam and an isotropic type-II OV.

Remarkably, since in both types of OV the phase behaves identically, performance of the polarization ellipses across the “full Poincare” region, particularly, the change in their azimuths when going around the C-point (C-area), occurs similarly in both cases. Therefore, for the ring-like area in Fig. 4d, all the regularities of the ellipse field formation, characteristic for a conventional field with a C-point, are fulfilled.

A remarkable peculiarity of the superposition with the type-II OV is that the central C-area is always polarized like the Gaussian component but at $\rho = a$ the handedness abruptly inverts. Practically, the abrupt change is impossible, and the handedness inversion develops on a certain interval, which means that, in fact, an additional s-contour appears near the ring $\rho \lesssim a$ in the case of type-II OV. But, topologically, the polarization-ellipses' pattern within the ring is identical to that near a C-point. Accordingly, such a region can be characterized by a topological index and a topological charge of the main phase, with the absolute value 1/2 (see, for example, Refs. [3,9]), as if a conventional C-point were at the field center.

Therefore, the resulting fields in the areas of the polarization modulation (circle $\rho < \rho_{\mathrm{sI}}$ in Fig. 4a, b, and the ring $a < \rho < \rho_{\mathrm{sII}}$ in Fig. 4c, d) are of practically identical topologies, except that, in the type-I case, the prevailing handedness is dictated by the Gaussian component field, whereas in the type-II case – by the handedness of the vortex contribution.

## 5. Formation of the type-II optical vortices

In [27,28], the formation of type-II OVs in optical domain was demonstrated in the process of smooth-wave passage through a transparent substrate on which metal nanodisks are located, as well as in phonon-polariton fields around nanoholes in a polariton plate. Particular attention should be paid to the method presented there for the formation of type-II OVs during the interference of a point dipole source and a plane wave, where the sign of the vortex is determined by the phase interrelation between these waves. At the same time, all the proposed solutions, despite their effectiveness, have a significant drawback: the difficulties that arise during their practical implementation, associated with the "micro" or even "nano" scale of the field configurations obtained. Besides, the fields generated and described in [27,28], are of essentially 2D nature inherent in evanescent optical fields, and cannot be applied in the framework of conventional optics dealing with the propagating beam-like fields.

In this context, the question arises: Is it possible to find other methods for forming fields with type-II OVs with scales that significantly exceed the subwavelength scale of the wave formations described in [27,28]? In principle, the positive answer to this question is formulated at the initial paragraphs of Section 3 (Eqs. (11), (12)). Now we discuss the relevant problems in more detail, keeping in mind the standard approaches to the structured light-field generation employing the SLM and synthesized holograms.

### *5.1. Principles of the type-II OV generation with SLM*

Naturally, this process starts with the registration of the amplitude $A(x,y)$ or intensity $A^2(x,y)$ profile of the "source" type-I field (11). At the second stage, the SLM is used for creation of the transparency with the variable transmission coefficient $\propto [A(x,y)]^{-2}$ with the "excluded" areas near the OV cores; being illuminated by the source wave, such a transparency produces the sought type-II field of the form (12) (or, which is more exactly, (17)) at the output plane.

Important precautions should be made in respect to the assignment and implementation of the hole areas described by the multiplier $[1 - Q_a(x - x_l, y - y_l)]$ in Eq. (17). The most direct way to determine the hole positions is based on the reaching a certain low-intensity level in the source field, which may lead to possible confusing a vortex with a deep minimum of intensity. However, such "erroneous" dark regions in the output field would not affect its phase distribution and especially the overall field behavior.

In the simplest case, a single type-II OV can be obtained from a conventional type-I OV-beam passed through a transparency whose transmittance drops sharply away from its center, while the central near-axis area is screened to form an excluded area (hole trace). The obvious drawback of this method is that creating small holes may be rather difficult, if not impossible, but, anyway, the subwavelength hole size is not a critical requirement. On the other hand, promising ways for the formation of type-II OVs may be based on the traditional techniques [10,29,30] using synthesized holograms, while the hologram center is “screened”, thus producing an imitation of a hole.

### *5.2. Structure and fabrication of the synthesized holograms for type-II field generation*

SLM can be employed also for creation of a synthesized (computer-generated) hologram – a special optical element where the desired optical field is formed in the course of diffraction. Other approaches, e.g., based on photolithographic hologram formation, are also relevant [45]. Regardless of the specific method of realization, the main problem in forming a corresponding mask is the fact that the visibility of the interference pattern with a smooth reference beam (plane wave or a Gaussian beam) varies greatly over the hologram area. The visibility variations are explained by the highly variative intensity of the type-II OV, especially, near the “intrinsic” boundary $\rho = a$ (see, for example, Eq. (13)). In the intensity distribution of the field (13),

$$I_V(\rho) = [1 - Q_a(\rho)]\left(\frac{|C_a|}{\rho}\right)^2, \tag{26}$$

the maximum intensity $I_{Vmax} = (|C_a|/a)^2$ is realized at $\rho = a$ (cf. Eqs. (15), (16)); inside the “excluded” area the intensity equals to zero (see Fig. 4c).

Let the reference wave be the plane wave with the complex amplitude distribution

$$U_p(x, y) = C_2 e^{ikx\sin\theta} \tag{27}$$

(we suppose that the wave direction vector lies in the (*x*, *z*) plane and makes an angle $\theta$ with the *z*-axis), and let there be no additional phase difference, unrelated with the structure of interfering waves ($C_2$ and $C_a$ are real quantities). Then, the intensity of the interference field obtained by superposition of the fields (13) and (27) is described by equation

$$I(x, y) = C_2^2 + \left[1 - Q_a\left(\sqrt{x^2 + y^2}\right)\right]\frac{C_a^2}{x^2 + y^2}$$

$$+2\left[1 - Q_a\left(\sqrt{x^2 + y^2}\right)\right]\frac{C_2 C_a}{\sqrt{x^2 + y^2}}\cos\left[kx\sin\theta - \arctan\left(\frac{y}{x}\right)\right]. \tag{28}$$

In particular, in the hole area $\rho < a$,

$$I(x, y) \equiv I_{hole} = C_2^2$$

while at the hole boundary $\rho \gtrsim a$, where the type-II field intensity reaches its maximum (26),

$$I(x, y) = C_2^2 + \frac{C_a^2}{a^2} + 2\frac{C_2 C_a}{a}\cos\left[kx\sin\theta - \arctan\left(\frac{y}{x}\right)\right]. \tag{29}$$

Hence, the maximum and minimum intensity of the interference pattern, achievable at the boundary $\rho = a$, can be obtained if the cosine in Eq. (29) equals to ±1:

$$I_{max} = \left(C_2 + \frac{C_a}{a}\right)^2, \quad I_{min} = \left(C_2 - \frac{C_a}{a}\right)^2. \tag{30}$$

As is well known (see, for example, Ref. [44]), the best diffraction efficiency is realized if the contrast of the hologram interference fringes equals to 1. Since this cannot be realized over the entire field area, we ensure this condition at the highest-OV-intensity location, at the hole boundary where the intensity is characterized by Eqs. (29), (30). There, the maximum-contrast requirement is expressed in the form

$$\frac{I_{max} - I_{min}}{I_{max} + I_{min}} = 1,$$

which entails $I_{min} = 0$, i.e. $C_a = aC_2$. Accordingly, if we accept this condition, the "optimal" intensity distribution of the hologram field, following from Eq. (28), can be expressed in the form

$$I(x,y) = C_2^2 \left\{ 1 + \left[ 1 - Q_a\left(\sqrt{x^2+y^2}\right)\right] \frac{a^2}{x^2+y^2} \right.$$

$$\left. +2\left[1 - Q_a\left(\sqrt{x^2+y^2}\right)\right] \frac{a}{\sqrt{x^2+y^2}} \cos\left[kx\sin\theta - \arctan\left(\frac{y}{x}\right)\right] \right\}. \quad (31)$$

Remarkably, even this optimal interference pattern is, generally, highly inhomogeneous so that the linearity conditions for the process of hologram recording are practically unavailable. Note also that, although in Eqs. (13) and (26) – (31), the unit absolute topological charge of the type-II OV is implied, one can easily reproduce the same reasoning for any integer topological charge $S$, but the resulting hologram will be still more inhomogeneous. For this reason, we restrict the consideration to the case of $|S| = |\sigma| = 1$; if necessary, higher-order OVs can be obtained in diffraction fields emerging in higher orders of diffraction at the hologram [8,46,47].

The patterns of the synthesized holographic masks (positive and negative) are presented in Fig. 5. The results of their practical application will be presented later in the Section 7.

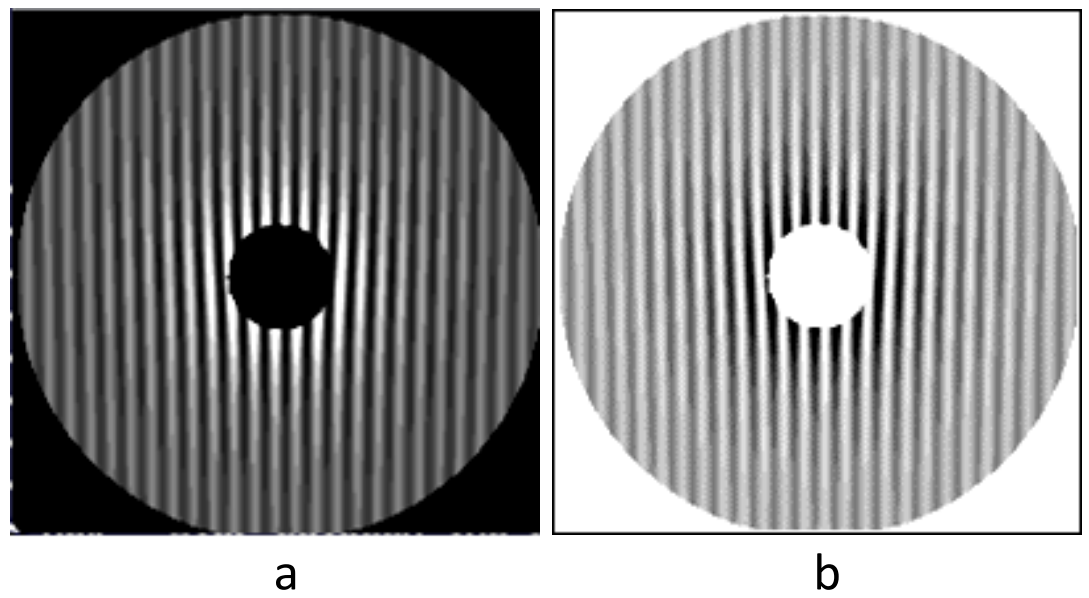

a b

Fig. 5. View of the holographic grating (mask) for the type-II OV generation in the diffracted field: (a) positive; (b) negative.

## 6. Free-space transformation of the type-II OVs

As was shown in Section 2, propagating type-II OVs can only exist in presence of the 3D holes impenetrable for the electromagnetic field and thus realizing the multiply connected space for optical waves. At the same time, one can easily check that the function describing an "ideal" scalar type-II OV with an arbitrary integer topological charge,

$$V_0(x,y,z) = \frac{C_0}{(x - i\sigma y)^{|S|}} e^{ikz} = \frac{C_0}{\rho^{|S|}} e^{i(S\varphi + kz)}, \quad \sigma = \mathrm{sgn}(S), \quad (32)$$

satisfies the 3D Helmholtz equation $\nabla^2 V_0 + k^2 V_0 = 0$, and thus represents a theoretically possible propagating wave field, although physically unrealizable in the exact form because of the singularity. In Sections 3 – 5, the physical structures modeling the field (32) in a local plane were considered, which enabled analysis of the main structural and topological characteristic of the type-II fields. A specific feature of these models is that they are "regularized" by the multiplier $[1 - Q_a(x, y)]$ (see Eqs. (13), (14), (17)) which plays a "dual" role. On the one hand, it models a 2D "trace" of the hole creating a multiply connected space, on the other hand it eliminates the singularity and makes the corresponding field physically realizable. However, its formal validity is not supported by the physical consistence: the region, physically unavailable for the field, is described as a deliberately assigned area of zero amplitude. Accordingly, the hole region appears to be available for other fields of the same nature (as we see in the superposition considered in Section 5.2). Moreover, the hole boundary, distinctly determined in the model plane, will be inevitably blurred, and the field will penetrate into the hole region, once we consider minor deviations from the initial transverse plane in the longitudinal *z*-direction.

In other words, if there are no special material structures realizing the "true" holes and producing the multiply connected space for the electromagnetic field, the 2D type-II configurations, described in Sections 3 – 5, will be inevitably destroyed upon propagation. In essence, from the perspectives of 3D propagation in free space, the type-II OVs, formed in a local transverse plane according to recommendations of the above Sections, are identical to the usual type-I OV. This is clearly seen from a comparison of Figs. 4b and 4c: each of the vortex amplitude distributions (depicted by blue and cyan colors) has zero value at the axis ($\rho = 0$); with growing off-axial distance $\rho$, these increase, reach a maximum, and, finally, decay to zero at the transverse infinity. The difference is concluded in the details of this common behavior: in the type-I field (Fig. 4b), the amplitude grows linearly with $\rho$ (in the type-II field, Fig. 4c, the growth occurs in a stepwise manner at $\rho = a$), reaches a smooth maximum (in Fig. 4c, the maximum is sharp), and slowly decays afterwards (in Fig. 4c, the decay is much more rapid). Additionally, the whole amplitude profile of the type-I OV is much wider than that of the smooth component while for the type-II OV, the situation is opposite. However, these discrepancies do not affect the principles of the light field transformations during its propagation along the *z*-direction (except that the sharp details of the type-II field profile make the self-diffraction phenomena more expressive). That is why the details of the free-space propagation of such type-II OV structures, initially prepared in a local transverse plane, are of interest, and will be considered below.

### *6.1. Analytical studies*

Now we consider the transformation of a circularly symmetric type-II OV. In the paraxial approximation, the propagation-induced transformation of the complex amplitude $V(\rho, \varphi, 0)$, defined in the input plane *z* = 0, can be described by the Fresnel-Kirchhoff integral [48–53], which in the cylindrical coordinates obtains the form [53]

$$V(r, \phi, z) = \frac{k}{2\pi i z} e^{ikz} \exp\left(\frac{ik}{2z} r^2\right) \iint_\Omega V(\rho, \varphi, 0) \exp\left(\frac{ik}{2z}\rho^2\right) \times \exp\left[-i\frac{k\rho r}{z}\cos(\phi - \varphi)\right] \rho d\rho d\varphi \, . \tag{33}$$

where $r, \phi$ are the polar coordinates in the output plane *z*. For calculations, we take the initial function in the general type-II OV form (32). Additionally, we take into account that the central part of the field is screened like in Eqs. (13), (14); besides, in any practical situation, the transverse size of the of the observable field is limited by a certain external boundary whose radius is assigned as $\rho = b$. Then,

$$V(r,\phi,z) = \frac{k}{2\pi i z} e^{i(S\phi+kz)} \exp\left(\frac{ik}{2z}r^2\right) \int_a^b \frac{C_0}{\rho^{|S|-1}} \exp\left(\frac{ik}{2z}\rho^2\right) d\rho$$

$$\times \int_{-\pi}^{\pi} \exp\left[iS\varphi - i\frac{k\rho r}{z}\cos(\phi-\varphi)\right] d\varphi$$

$$= i^{S-1}\frac{k}{z} e^{i(S\phi+kz)} \exp\left(\frac{ik}{2z}r^2\right) \int_a^b \frac{C_0}{\rho^{|S|-1}} \exp\left(\frac{ik}{2z}\rho^2\right) J_{|S|}\left(\frac{k\rho r}{z}\right) d\rho \qquad (34)$$

where $J_{|S|}(h)$ denotes the Bessel function of the 1st kind [54].

Let us first consider the asymptotic behavior at $z \to \infty$ (far-field diffraction); in addition to the analytical convenience, the far-field patterns are realized, e.g., at the focal plane of a lens and are used in many practical situations [48–51]. At this condition, the exponential term inside the integral may be replaced by the unity, and the whole integral in (34) can be calculated in a closed form [54], which ultimately gives

$$\left[V(r,\phi,z)\exp\left(\frac{ik}{2z}r^2\right)\frac{z}{k}e^{-ikz}\right]_{z\to\infty} \equiv \tilde{V}(r,\phi,p)$$

$$= i^{S-1}e^{iS\phi}\frac{C_0}{kp}\left[\frac{1}{a^{|S|-1}}J_{|S|-1}(kpa) - \frac{1}{b^{|S|-1}}J_{|S|-1}(kpb)\right] \qquad (35)$$

where $p = r/z$ is the far-field angular variable. Generally, it is the function $\tilde{V}(r,\phi,p)$ that provides a consistent far-field characterization of the beam profile. Especially, near the axis (more specifically, upon the condition $kpb \ll 1$), the approximate expression of the Bessel function [54]

$$J_{|S|-1}(h) \approx \frac{1}{(|S|-1)!}\left(\frac{1}{2}h\right)^{|S|-1}\left[1 - \frac{1}{|S|}\left(\frac{1}{2}h\right)^2\right]$$

immediately yields

$$\tilde{V}(r,\phi,p) = \frac{1}{2}i^{S-1}e^{iS\phi}\left(\frac{1}{2}kp\right)^{|S|}(b^2-a^2), \qquad (36)$$

or, for the "generic" case $|S| = 1$,

$$\tilde{V}(r,\phi,p) = \frac{1}{4}\sigma e^{i\sigma\phi}kp(b^2-a^2). \qquad (37)$$

Remarkably (and expectedly), the vortex structure of the field, expressed by the multiplier $e^{iS\phi}$, is preserved while its type-II feature is completely lost: in the far field, a conventional type-I OV is formed with the power-law near-axis amplitude growth.

### *6.2. Numerical investigation*

For better understanding the regularities of the type-II OV transformations occurring during the free propagation, it is necessary to examine how its structure evolves with growing but finite propagation distance $z$; in other words, one needs to trace the beam behavior in the Fresnel-diffraction region [53]. This can be performed based on Eq. (34); however, for finite $z$, there is no explicit expression for the evolving field amplitude. The only meaningful conclusion which can be derived from the general expression is that the propagating field preserves the circular symmetry of the intensity distribution, together with the helical phase expressed by the phase factor $e^{iS\phi}$; other details of the field evolution can be revealed via the numerical analysis.

The results of computer modeling are presented in Figs. 6 and 7, which show transformations of the type-II OV field intensity and phase, depending on the propagation distance $z$ and the "hole trace" diameter in the initial plane $2a$.

Fig. 6 illustrates the field evolution in the Fresnel zone. It shows the gradual transformation of the initial type-II structure into a conventional type-I OV. At the propagation distance $z \gtrsim 1000$ mm, the field pattern qualitatively does not differ from the pattern inherent in the "classic" Laguerre-Gaussian or Bessel beam [12]. Some curvature of equiphase lines can be explained by the quadratic phase addition originating from the spherical wave front of the diverging beam.

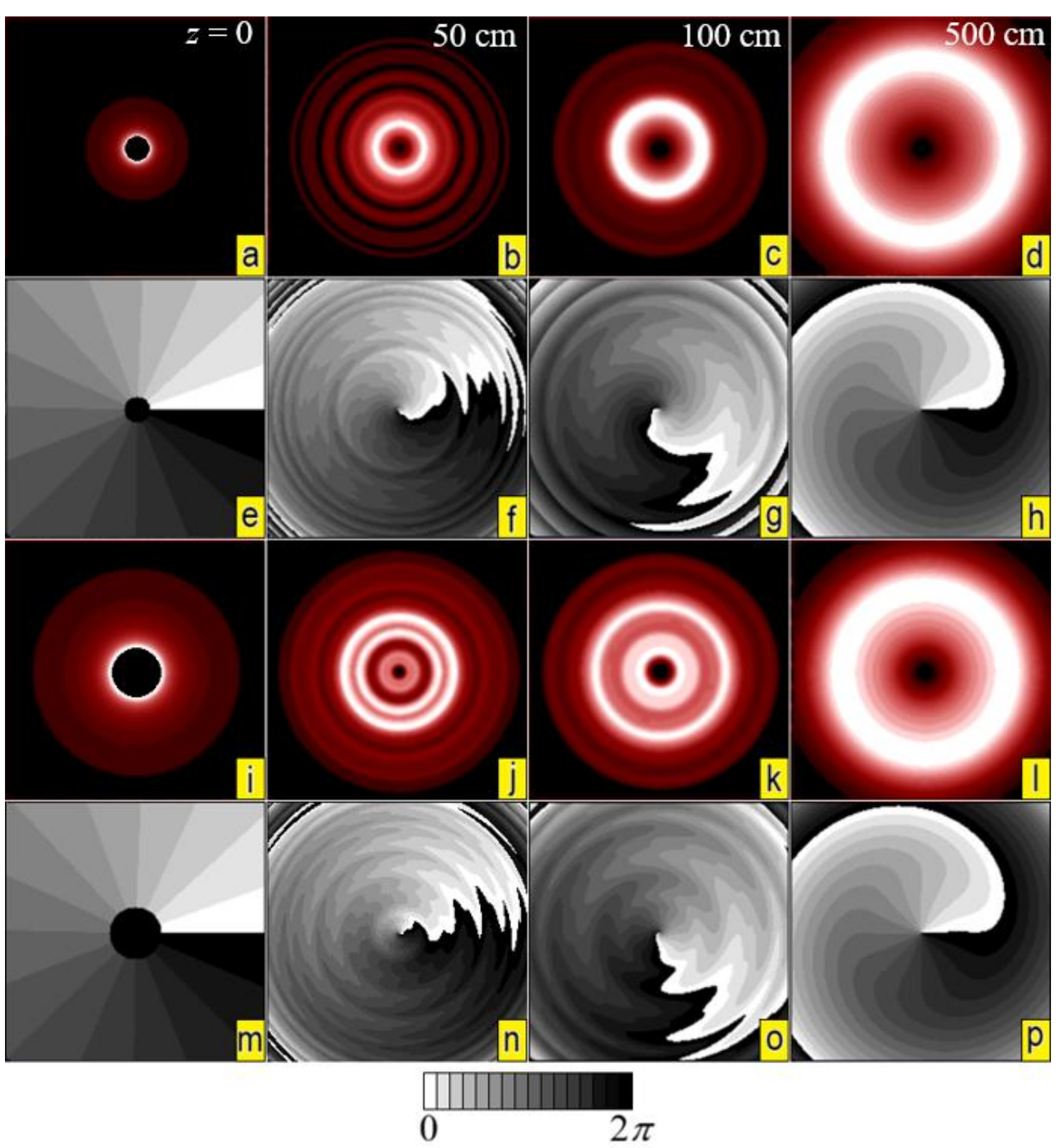

Fig. 6. Results of the computer modeling of the type-II OV propagation according to Eq. (34), with the topological charge $|S| = 1$. The radiation wavelength is 0.63 μm (He-Ne laser), the diameter of the external diaphragm bounding the field in the initial plane (plane of generation) is $2b = 4$ mm, the hole diameter is $2a_1 = 0.4$ mm in panels (a) – (h) (two upper rows) and $2a_2 = 0.8$ mm in panels (i) – (p). Images (a) – (d) and (i) – (l) (1st and 3rd rows) show the intensity distributions, (e) – (h) and (m) – (p) (2nd and 4th rows) – phase distributions. The left column illustrates the initial distributions in the plane of the OV generation $z = 0$, described by Eqs. (12), (13) with $\sigma = 1$; further columns correspond to $z = 50$ cm, $z = 100$ cm, and $z = 500$ cm.

Fig. 7 shows the results of modeling the type-II field propagation in the near-field zone. As is seen, a type-II OV transforms fairly quickly into a type-I vortex. At the same time, a region of relatively low intensity persists at a significant distance from the generation plane, with an absolute (zero) minimum at the beam axis. The results of modeling confirm the unambiguous conclusion that type-II OVs, artificially created in a single plane with modeling a hole as a zero-amplitude area (described by expression (13) and its analogs), are non-generic structures with respect to the 3D propagation in free space. They exist only in the plane of their generation or in the image plane if, in the process of their propagation, the wave passes through a projective optical system.

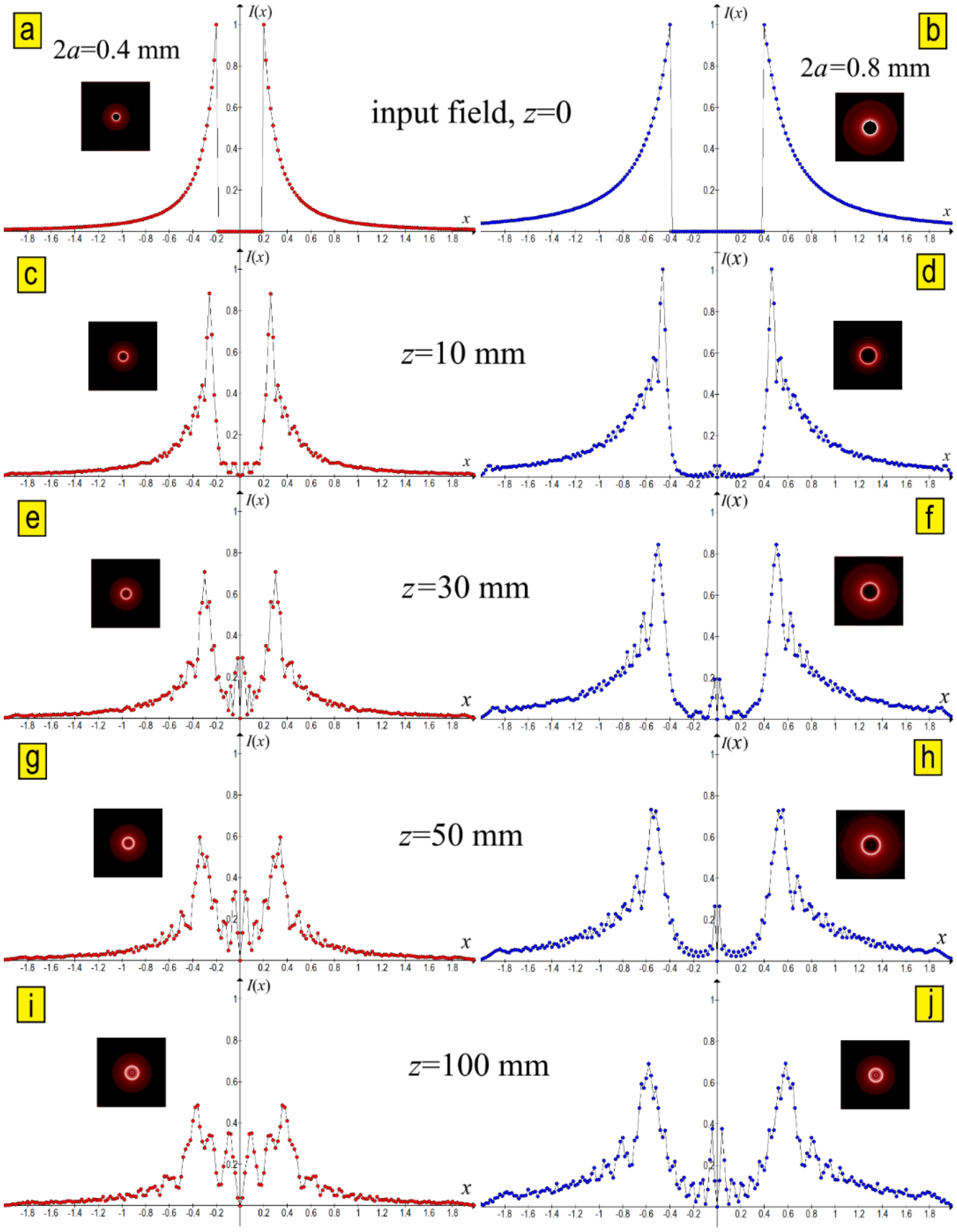

Fig. 7. Evolution of the transverse intensity profile upon the near-field propagation of the type-II OV with the initial distribution at the plane of generation given by Eqs. (12), (13) with $\sigma = 1$. The radiation wavelength is 0.63 μm, the external diaphragm diameter in the initial plane is $2b$ = 4 mm, the hole diameter is $2a_1$ = 0.4 mm (left column) and $2a_2$ = 0.8 mm (right column). Additionally, the insets in each panel show the corresponding 2D intensity profiles (all the distributions are circularly symmetric, the main images present their sections corresponding to $y = 0$). The intensity values are normalized with respect to the maximum value that is observed at the hole boundary $x = a$ in the generation plane. Distance of propagation: (a), (b) $z = 0$; (c), (d) $z = 10$ mm, (e), (f) $z = 30$ mm, (g), (h) $z = 50$ mm, (i), (j) $z = 100$ mm.

However, it is interesting to evaluate the field's ability to self-preserve and its resistance to inevitable distortions caused by the free propagation. This quality can be characterized by the distance $z_0$ such that for $z < z_0$, the type-II OV intensity distribution may be considered the same as in the initial plane (the "shadow approximation" [53] is valid). Actually, this distance is determined by the diffraction at a circular aperture of the radius $a$, which means that the validity limit of the shadow approximation can be expressed as [53]

$$z_0 \le \frac{a^2}{\lambda}\,. \tag{38}$$

If the optical system produces an image of the initial plane, this distance determines the longitudinal size of a certain dark volume surrounded by the narrow region of a high intensity and, consequently, of high OAM density (see Eqs. (15), (16)). The corresponding field configuration (Fig. 8) can be likened to a specific 3D light trap – a so called "bottle beam" (see, e.g., Refs. [55–58]). The bottle length can be estimated as

$$L = 2z_0 = \frac{2a^2}{\lambda}\,. \tag{39}$$

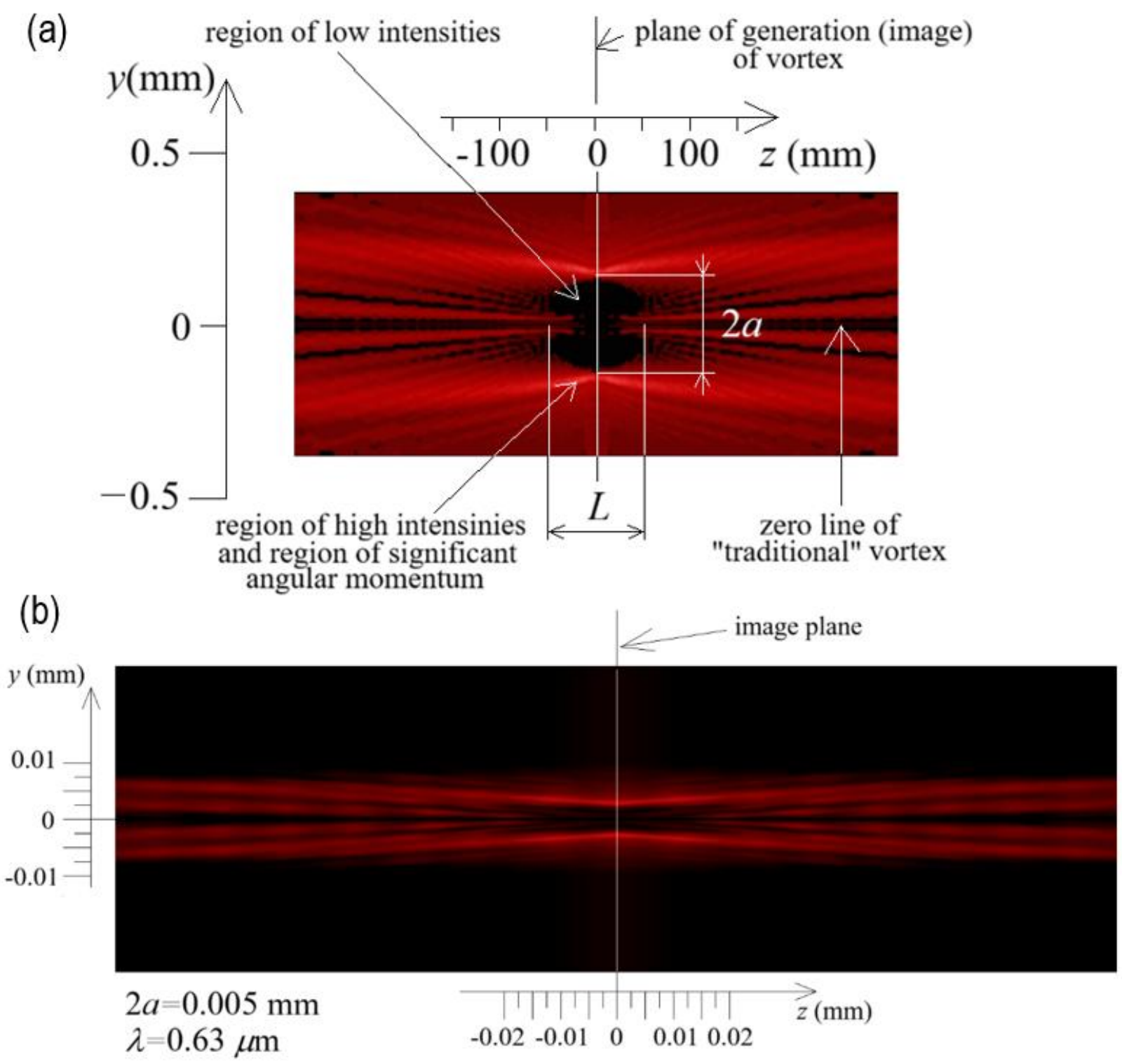

Fig. 8. The longitudinal section of the propagating type-II OV field near the image plane ("bottle" structure). The radiation wavelength $\lambda = 0.63\ \mu$m, the hole diameter $2a$ equals to (a) 0.4 mm; (b) 0.005 mm.

Accordingly, the ratio $\beta$ between the width of the trap and its length can be estimated from the relation

$$\beta = \frac{\text{width}}{\text{length}} = \frac{\lambda}{a}\,. \tag{40}$$

In Fig. 8b, a longitudinal cross-section of the field near the image plane is shown for the type-II OV hole diameter $2a = 0.0025$ mm and a wavelength $\lambda = 0.63$ μm. According to relation (40), the trap length in this case exceeds its width by approximately 4 times.

## 7. Experimental verification

Experimental studies of the type-II field generation and propagation were performed with the arrangement depicted in Fig. 9. In this scheme, a light source is the He-Ne laser 1 ($\lambda = 0.63$ μm). The output beam of the laser forms the object and reference channels using a beam splitter 2. A vortex hologram 3, fabricated according to the methodology described in Section 5 and forming a type-II OV, is placed in the object channel. An assigned type-II OV structure is formed using the telescopic system 4, 5. As a result, the "ideal" type-II OV field of the form (13) is formed in the image plane. In this plane ("generation plane"), the initial measurements were taken. Afterwards, the elements 6, 8, and 9 were moved to various positions such that the total propagation distance from the generation plane is $z = l_1 + l_2 + l_3$ . The resulting transformed type-II fields, as well as the patterns of their interference with the reference wave, are formed at the input plane of the CCD camera, where the fields' intensity distributions are registered, transmitted to a PC (not shown), processed and analyzed.

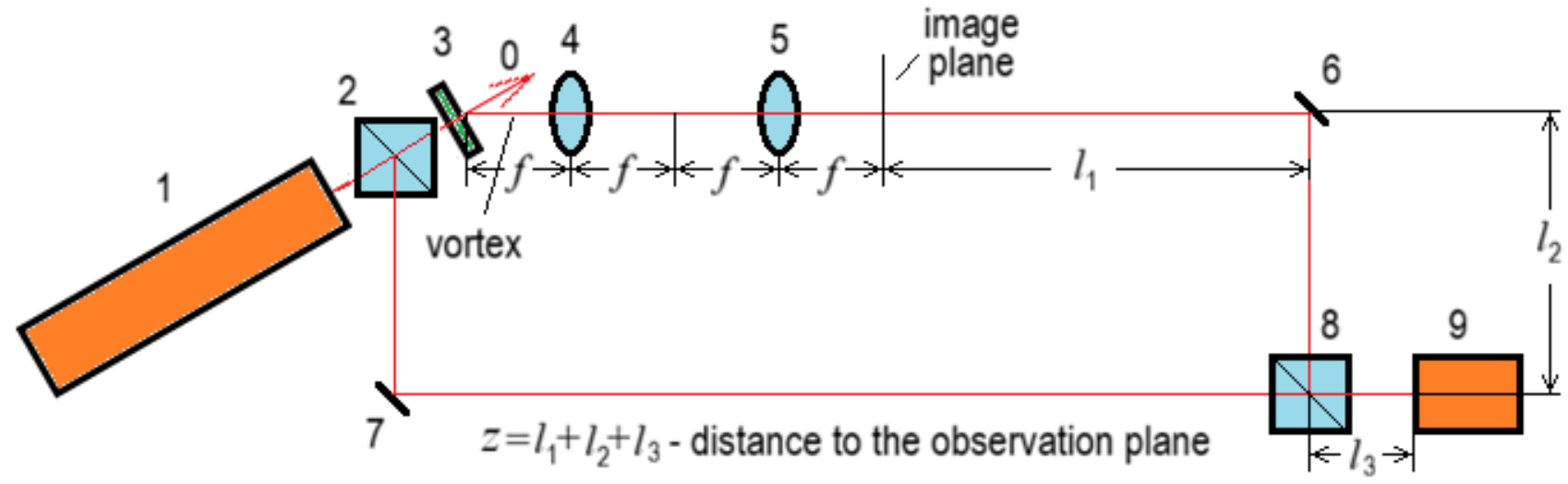

Fig. 9. Arrangement for the type-II OV investigation: (1) He-Ne laser; (2), (8) beam splitters; (3) hologram of the sort presented in Fig. 5; (4), (5) telescopic lens system; (6), (7) mirrors; (9) CCD-camera.

The results are presented in Figs. 10 and 11. As is seen in Fig. 10a, our holographic technique, indeed, produces a "classical" type-II OV field of Eq. (13) with the dark hole in the center and a sharp intensity maximum at its boundary. However, during propagation, this field gradually transforms into a conventional type-I OV, in full agreement to the theory of Section 6. In the near field (panels (b) and (c)), the bright intensity maximum is a bit blurred, and the dark hole shrinks: this behavior illustrate the "bottle beam" zone discussed numerically in Fig. 8. Panels (d) – (h) show the intensity distributions in the Fresnel zone of the beam propagation, where the main observable transformation is that the central "gap" is completely destroyed and replaced by the multi-ring diffraction structure predicted in Fig. 7i, j, and 6b, c, j, k.

Additionally, Fig. 11 illustrates the interference between the propagating type-II OV beam and the reference beam (see Fig. 9). As can be seen from Fig. 11b, the number of fringes above the dark hole region exceeds the number of fringes beneath this region by 1, which testifies the helical phase distribution and thus confirms the formation of a type-II OV with the topological charge 1 in the plane $z = 0$. When forming the interferogram in the Fresnel zone, the so-called "fringe zero" setting was used. The structure of the interferogram in Fig. 11d at a small distance from the field axis corresponds to the characteristic spiral interferogram formed by a conventional type-I OV with the same charge. The nontrivial shape of the interferogram fringes at the beam periphery can be explained by the fact that, as can be seen from Fig. 11c, the OV core is surrounded by "diffraction rings" with accompanying $\pi$-jumps of the phase. A resulting phase distribution of the field presented in Fig. 11d is thus similar to the phase distribution obtained from numerical simulation for the comparable conditions (Figs. 6g, 6o).

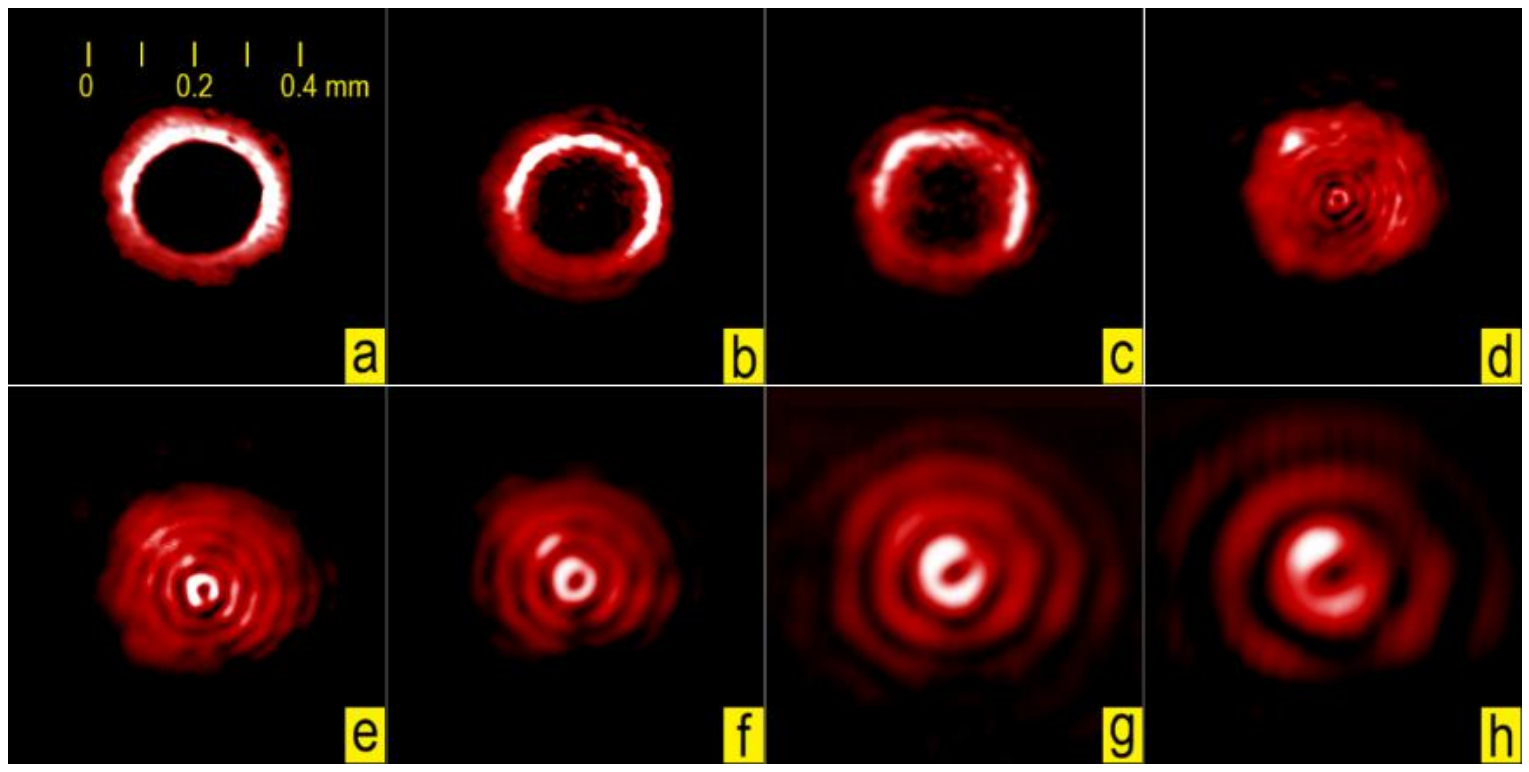

Fig. 10. The intensity profile transformations during the type-II OV propagation: (a) plane of generation $z = 0$; (b) $z = 5$ cm; (c) $z = 10$ cm; (d) $z = 25$ cm; (e) $z = 40$ cm; (f) $z = 50$ cm; (g) $z = 80$ cm; (h) $z = 120$ cm. The hole diameter in the initial plane (panel (a)) is $2a = 0.25$ mm.

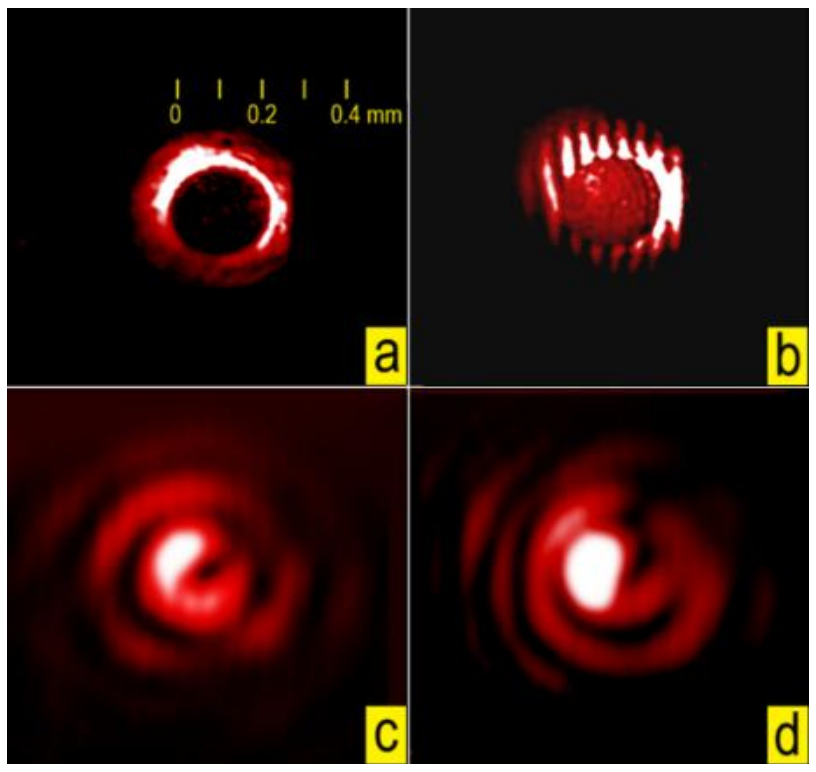

Fig. 11. Intensity distributions (a), (c) and patterns of interference with the reference wave (see Fig. 9) (b), (d) for the type-II OV: (a), (b) in the initial (generation) plane, $z = 0$; (c), (d) in the Fresnel diffraction zone, $z = 120$ cm (cf. Fig. 10h).

In general, our preliminary experimental results show a fair qualitative agreement with the theoretical predictions of Section 6. In particular, these testify that:

- A model of the type-II OV in the form (13) can be created in a single plane (generation plane), with the help of the traditional techniques using the computer-synthesized holograms [29,30];
- On the free propagation, this type-II OV transforms into the conventional type-I OV. Remarkably, in the near-field stage of its propagation, a 3D dark volume is formed, which can be used as an optical "hollow" trap of the bottle-beam [55–58] type.

## 8. Discussion and conclusion

In this paper, we have considered possibilities for the optical 3D generalization of the type-II OVs, recently introduced as the 2D objects [27,28]. The undertaken analysis as well as numerical simulations and preliminary experiments have led us to the following conclusions.

The propagating (3D) type-II OVs can only be realized in the multiply connected space, or in presence of material inclusions which model the multiply-connected space for the optical waves, e.g., creating the "holes" – spatial regions impenetrable for electromagnetic field. Such material inclusions can be fabricated, for example, based on the perfect metal model [32], and enable a diffraction-free and lossless propagation of the type-II OVs, at least approximately in the paraxial regime (Section 2).

However, in the usual single-connected free space, the stable type-II OVs are impossible. Only their cross sections can be realized in a single plane (plane of generation) where the trace is modeled as a zero-amplitude region near the OV core. In this plane, the type-II OVs can be considered as specific vortex-like singular formations, which participate in non-trivial and instructive interrelations, forming the singular networks of the optical field. Since the phase topology in the type-II OVs is the same as of the type-I OVs, the regularities of the singular networks in the type-II fields mimic those existing in the usual speckle fields with the conventional type-I OVs. Such "planar" type-II fields can interact with the usual optical fields, represented by their cross sections taken in the same single plane; they can be created independently of the field polarization (within the generation plane), and the scalar type-II OVs can be considered and realized. On the other hand, inhomogeneously polarized type-II OV fields, including the fields with polarization singularities, may exist (Sections 3, 4).

There are various ways for the physical realization of such planar type-II OVs, including the source-beam passage through a specially prepared transparency, application of an SLM, or employment of the traditional techniques based on the synthesized (computer-generated) holograms [29,30].

However, according to the above-mentioned instability of the type-II OV in free space, any deviation from the plane of generation leads to its destruction. Upon free propagation, a type-II OV fairly quickly transforms into a traditional type-I vortex characterized by the power-law amplitude growth with the off-core distance. The transformation rate is determined by the hole diameter and the radiation wavelength (see Eq. (38)).

It should be noted that the instability of the type-II OVs upon propagation may be useful for the formation of bottle beam optical traps [55–58] with unique characteristics. As previously stated, unlike a conventional type-I OV, in which the near-core intensity decreases according to a power law on approaching the beam center, the intensity of a type-II OV increases sharply near the hole boundary. For micro-objects which tend to locate near the field minima, an effective "hollow trap" is formed with strong "side walls". Upon the near-field propagation, the intensity "jump" becomes smoother and, anyway, the radiation penetrates into the beam center, forming the "front" and "back" walls. Actually, there appears a "bottle" near the generation plane: a quasi-hollow low-intensity volume surrounded by high-intensity walls. Remarkably, the "side walls" concentrate not only the light energy but also the OAM, which opens efficient ways for the trapped object's manipulations. The latter feature is common, e.g., with the "perfect vortex beams" possessing a thin bright-ring intensity profile [59], but only the type-II OVs form a 3D bottle trap additionally.

A trap of this type can be intentionally formed in a specific plane, for example, using an optical system that forms an image of the type-II OV generation plane. The transverse trap size is determined by the hole diameter $2a$ while the longitudinal size can be estimated as the longitudinal limit of the shadow approximation $2z_0$ (see Eq. (38)).

**Funding.** This work was supported by the National Research Foundation of Ukraine under grant № 2025.06/0086.

**Disclosures.** The author declares no conflicts of interest.

**Data availability.** Data underlying the results presented in this paper are not publicly available at this time but may be obtained from the authors upon reasonable request.